\documentclass[aip,reprint]{revtex4-1}
\usepackage[space]{grffile}
\usepackage{bm,graphicx,graphics,amsmath,amssymb,bm,epsfig,color}
\usepackage{euscript,tabularx}
\usepackage{longtable}
\usepackage{float}

\begin{document}

\title{Measurement of the intrinsic damping constant in individual
  nanodisks of YIG and YIG{\bf \textbar}Pt}

\author{C. Hahn}
\affiliation{Service de Physique de l'\'Etat Condens\'e (CNRS URA
  2464), CEA Saclay, 91191 Gif-sur-Yvette, France}

\author{V.V. Naletov}
\affiliation{Service de Physique de l'\'Etat Condens\'e (CNRS URA
  2464), CEA Saclay, 91191 Gif-sur-Yvette, France}
\affiliation{Unit\'e Mixte de Physique CNRS/Thales and Universit\'e
  Paris Sud 11, 1 av. Fresnel, 91767 Palaiseau, France}
\affiliation{Institute of Physics, Kazan Federal University, Kazan
    420008, Russian Federation}

\author{G. de Loubens}
\affiliation{Service de Physique de l'\'Etat Condens\'e (CNRS URA
  2464), CEA Saclay, 91191 Gif-sur-Yvette, France}

\author{O. Klein}
\affiliation{Service de Physique de l'\'Etat Condens\'e (CNRS URA
  2464), CEA Saclay, 91191 Gif-sur-Yvette, France} \email[Corresponding author:]{ oklein@cea.fr}

\author{O. d'Allivy Kelly}
\affiliation{Unit\'e Mixte de Physique CNRS/Thales and Universit\'e
  Paris Sud 11, 1 av. Fresnel, 91767 Palaiseau, France}

\author{A. Anane}
\affiliation{Unit\'e Mixte de Physique CNRS/Thales and Universit\'e
  Paris Sud 11, 1 av. Fresnel, 91767 Palaiseau, France}

\author{R. Bernard}
\affiliation{Unit\'e Mixte de Physique CNRS/Thales and Universit\'e
  Paris Sud 11, 1 av. Fresnel, 91767 Palaiseau, France}

\author{E. Jacquet}
\affiliation{Unit\'e Mixte de Physique CNRS/Thales and Universit\'e
  Paris Sud 11, 1 av. Fresnel, 91767 Palaiseau, France}

\author{P. Bortolotti}
\affiliation{Unit\'e Mixte de Physique CNRS/Thales and Universit\'e
  Paris Sud 11, 1 av. Fresnel, 91767 Palaiseau, France}

\author{V. Cros}
\affiliation{Unit\'e Mixte de Physique CNRS/Thales and Universit\'e
  Paris Sud 11, 1 av. Fresnel, 91767 Palaiseau, France}

\author{J.L. Prieto} 
\affiliation{Instituto de Sistemas Optoelectr\'onicos y
  Microtecnolog\'{\i}a (UPM), Madrid 28040, Spain}

\author{M. Mu\~noz}
\affiliation{Instituto de Microelectr\'onica de Madrid (CNM, CSIC),
  Madrid 28760, Spain}

\date{\today}

\begin{abstract}
  We report on an experimental study on the spin-waves relaxation rate
  in two series of nanodisks of diameter $\phi=$300, 500 and 700~nm,
  patterned out of two systems: a 20~nm thick yttrium iron garnet
  (YIG) film grown by pulsed laser deposition either bare or covered
  by 13~nm of Pt. Using a magnetic resonance force microscope, we
  measure precisely the ferromagnetic resonance linewidth of each
  individual YIG and YIG{\textbar}Pt nanodisks. We find that the
  linewidth in the nanostructure is sensibly smaller than the one
  measured in the extended film. Analysis of the frequency dependence
  of the spectral linewidth indicates that the improvement is
  principally due to the suppression of the inhomogeneous part of the
  broadening due to geometrical confinement, suggesting that only the
  homogeneous broadening contributes to the linewidth of the
  nanostructure. For the bare YIG nano-disks, the broadening is
  associated to a damping constant $\alpha = 4 \cdot 10^{-4}$. A 3
  fold increase of the linewidth is observed for the series with Pt
  cap layer, attributed to the spin pumping effect. The measured
  enhancement allows to extract the spin mixing conductance found to
  be $G_{\uparrow \downarrow}= 1.55 \cdot 10^{14}~
  \Omega^{-1}\text{m}^{-2}$ for our YIG(20nm){\textbar}Pt interface,
  thus opening large opportunities for the design of YIG based
  nanostructures with optimized magnetic losses.
\end{abstract}

\maketitle


\begin{table}
  \caption{Magnetic parameters of the 20~nm thick YIG film.}
  \begin{ruledtabular}
    \begin{tabular}{c c c c c}
      $4 \pi M_s$ (G) &  $\alpha$  & $\Lambda_{ex}$ (nm) & 
      $\gamma$ (rad.s$^{-1}$.G$^{-1}$) \\ \hline \\
      $2.1 \cdot 10^{3}$ & $4 \cdot 10^{-4}$ & 15 & 
      $1.79\cdot 10^{7}$  \\
    \end{tabular}
  \end{ruledtabular}\label{tab:param}
\end{table}

Yttrium iron garnet (Y$_3$Fe$_5$O$_{12}$), commonly referred as YIG,
is the champion material for magneto-optical applications as it holds
the highest figure of merit in terms of low propagation loss. It is
widely used in high-end microwave and optical-communication devices
such as filters, tunable oscillators, or non-reciprocal devices. It is
also the material of choice for magnonics \cite{serga10}, which aims
at using spin-waves (SW) (or their quanta magnons) to carry and
process information. The development of this emerging field is
presently limited by the damping constant of SW. Recently it was
proposed that spin-current transfer generated by spin Hall effect from
an adjacent layer can partially or even fully compensate the intrinsic
losses of the traveling SW beyond the natural decay time. The
achievement of damping compensation by pure spin current in 1.3$\mu$m
thick YIG covered by Pt was reported by Kajiwara \textit{et al.}
\cite{kajiwara10}, although attempts to reproduce the results have so
far failed \cite{hahn13,kelly13}. Given that the spin-orbit torque is
purely interfacial in such hybrid system, it is crucial to work with
nanometer-thick films of epitaxial YIG. Primarly, because the
interface spin-current transfer scales inversely with the YIG
thickness \cite{xiao12}. Secondly, because it permits nano-patterning
of the YIG and thus the engineering of the spin-wave (SW) spectra
through spatial confinement \cite{jorzick02,loubens07,guo13}. To the
best of our knowledge, however, there is no report yet
\cite{Pirro2014} on the measurement of the dynamical properties on
submicron size nanostructure patterned out of YIG ultrathin films.

Benefiting from our recent progress in the growth of very high quality
YIG films by pulsed laser deposition (PLD) \cite{kelly13}, here we
will demonstrate that these ultra-thin YIG films (thicknesses ranging
from 20 to 4~nm) can be reliably nano-patterned into sub-micron size
nano-disks. In the following, we will perform a comparative study of
the linewidth on these nanostructures. We will analyze the different
contributions to the damping by separating the homogeneous from the
inhomogeneous broadening and quantify the spin-pumping contribution
when an adjacent metallic layer is added. We will also evaluate the
consequences of the damages produced by the lithographic process on
the dynamical response of these devices. It will be shown that the
different alterations produced by chemical solvent, heat treatment,
amorphization and redeposition of foreign elements during the
lithography, edge roughness etc...  do not produce any increase to the
linewidth, offering great hope for incorporation of these YIG films in
magnonics.

\begin{figure}
  \includegraphics[width=0.8\columnwidth]{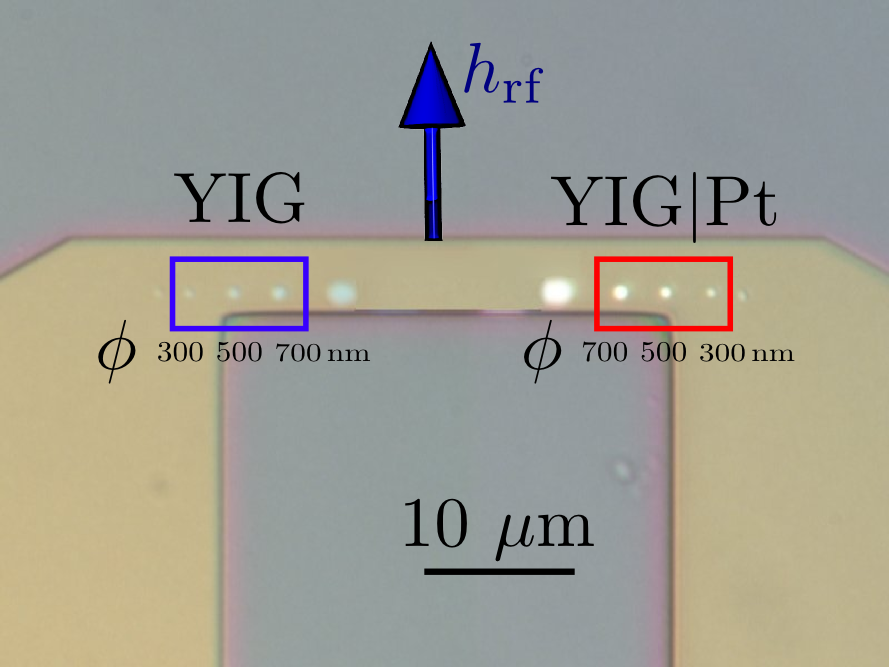}
  \caption{Optical image of the YIG disks, with (red) and
    without (blue) Pt on top, placed below a microwave antenna. The
    direction of the microwave magnetic field is indicated by a blue
    arrow. The external bias magnetic field is oriented
    perpendicularly to the surface. }
  \label{fig:mfm}
\end{figure}

The 20~nm thick YIG has been grown by PLD on a (111)
Gd$_3$Ga$_5$O$_{12}$ substrate following the preparation described in
Ref. \cite{kelly13}. A 13~nm thick Pt layer was then deposited by
sputtering on half of the YIG surface. One slab of $1 \times 5~$mm was
cut to perform standard magnetic hysteresis and ferromagnetic
resonance (FMR) measurements. The measured magnetic parameters of the
bare YIG film are summarized in Table \ref{tab:param}. On the
remaining piece, a series of YIG nanodisks has been subsquently
patterned using standard electron lithography and dry ion etching.
After the YIG lithography, an insulating layer of 50~nm SiO$_2$ was
deposited on the whole surface and a 150~nm thick and 5~$\mu$m wide
microwave Au-antenna was deposited on top. Fig.\,\ref{fig:mfm} shows
an optical image of the sample and the antenna pattern. The series of
decreasing diameters bare YIG nanodisks is placed on the left. The
spacing between the disks is 3 $\mu$m. The series of disks on the
right side mirrors the first one and has a 13~nm Pt layer on top. Here
we concentrate on the disks enclosed in the rectangular area, with
nominal diameter $\phi=$700, 500 and 300~nm.

\begin{table}
  \caption{
    Comparative table of the measured and predicted (analytical and
    SpinFlow simulations)\cite{Naletov2011} resonance values for the SW modes. These predictions are uncorrected by the additional
    stray field of the MRFM probe. The last row display the SWs
    eigen-values obtained by adjusting both the radius of the disks
    (respectively $\phi=700, 520$ and 380~nm) and the probe sample separation (respectively
    $h=3.0,2.0$ and 1.5$\mu$m). The SWs are labeled by their azimuthal and
    radial number $(\ell,n)$ \cite{Naletov2011}.}
  \begin{ruledtabular}
    \begin{tabular}{c | c c c c c }
$\phi$ (nm) & mode & $f_\text{exp}$ (GHz) & simu. & 
analyt. &  fit \\ \hline
700 & ${(00)}$ &  8.74 &  8.54 &  8.69 &  8.76 \\
700 & ${(01)}$ &  9.14 &  9.04 &  9.10 &  9.16 \\ 
700 & ${(02)}$ &  9.65 &  9.66 &  9.68 &  9.72 \\ \hline 
500 & ${(00)}$ &  9.10 &  8.73 &  8.90 &  9.05 \\
500 & ${(01)}$ &  9.72 &  9.51 &  9.63 &  9.69 \\ 
500 & ${(02)}$ & 10.52 & 10.51 & 10.72 & 10.65 \\ \hline
300 & ${(00)}$ &  9.58 &  9.19 &  9.47 &  9.51 \\ 
300 & ${(01)}$ & 10.57 & 10.66 & 11.25 & 10.63 \\
    \end{tabular}
  \end{ruledtabular}\label{tab:eigen}
\end{table}

To measure the FMR-spectra of the nanodisks buried under the microwave
antenna, we use a ferro-magnetic resonance force microscope (f-MRFM)
\cite{klein08}. It is based on measuring the deflection of a
MFM-cantilever with a magnetic Fe particle of about 800~nm diameter
affixed to the tip. The tip magnetic dipole moment senses the stray
field produced by the perpendicular component M$_z$ of the
magnetization of the magnetic nanodisks, which is modulated by the
exciting microwave power at the mechanical frequency of the
cantilever.

\begin{figure}[hb!]
  \includegraphics[width=1.\columnwidth]{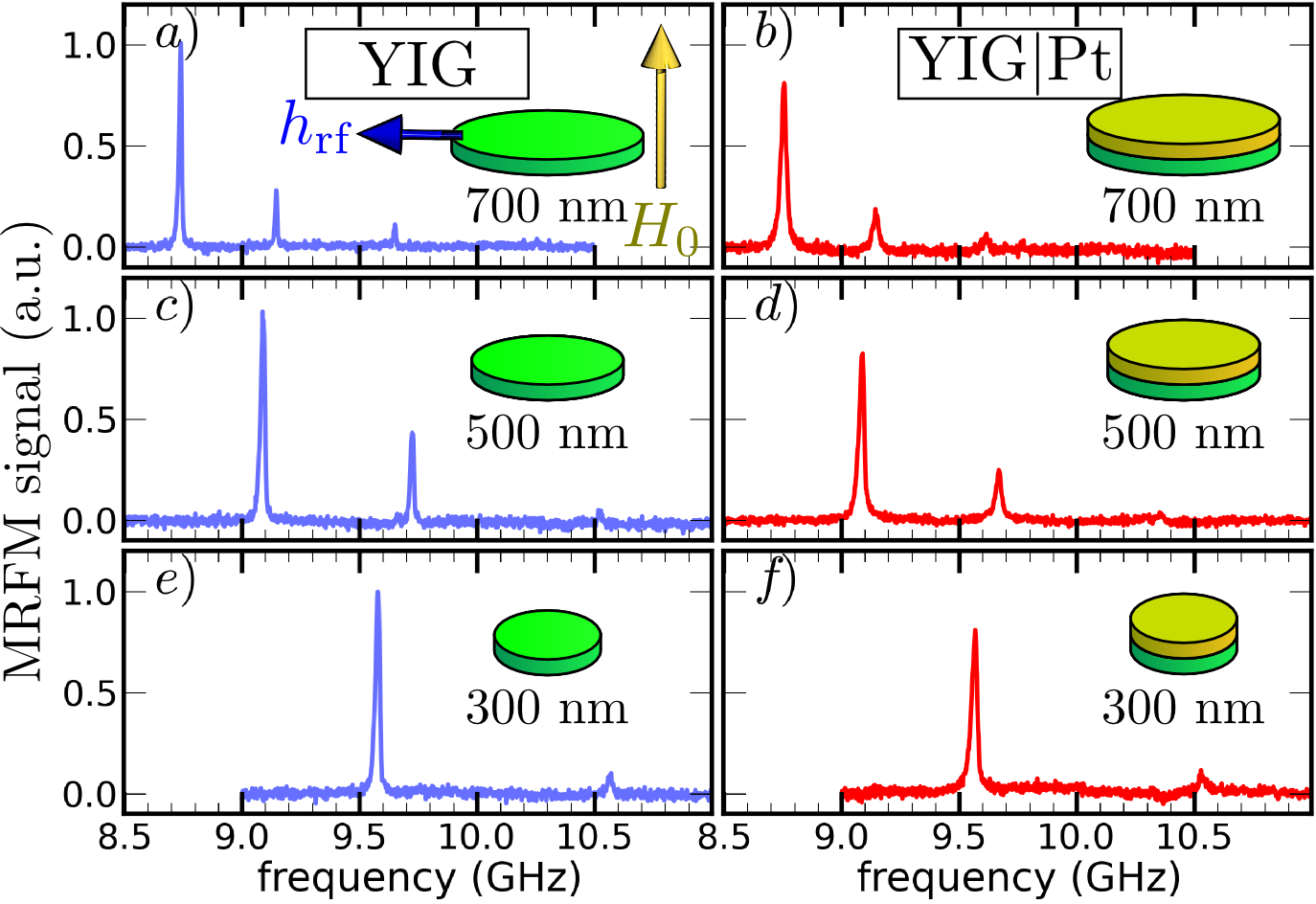}
  \caption{Mechanical-FMR spectra at $H_{0}$=4.99 kOe of the 700,
    500 and 300~nm diameter YIG disks arranged in rows by decreasing
    lateral size.  The spectra of the pure YIG disks are shown in the
    left column (blue), while the ones covered with 13~nm of Pt are
    shown in the right column (red).}
  \label{fig:compspec}
\end{figure}

In Fig.\,\ref{fig:compspec} we show f-MRFM spectra recorded for
different diameters (by row) on both the YIG and YIG{\textbar}Pt
nanodisks (by column). The spectra correspond to the spin-wave
eigenmodes of the disks biased by a perpendicular magnetic field
$H_{0}=4.99$~kOe. The largest peak at lowest frequency stems from the
lowest energy FMR-mode, the so-called uniform mode. The smaller peaks
at higher energy correspond to higher order modes. The splitting
corresponds to the quantization of the SW wavenumber $\propto n
\pi/\phi$ ($n$ being an integer) in the radial direction
\cite{klein08}. One can thus infer from the peak separation the
lateral size of the disk. Using the
literature\cite{sparks64,Damon1965} value for the YIG exchange length
$\Lambda_\text{ex}= 15$~nm, a fit of the peak separation leads to an
effective confinement of respectively 700, 520, 380~nm for our 3
disks, assuming total pinning at the disk edge (see Table II). This is
in very good agreement with nominal sizes targeted by the patterning.
Differences with the nominal value are due both to imperfection of the
lithographic process and the dipolar pinning condition
\cite{Guslienko2002}, which scales as the aspect ratio. Confining a
spin wave in a smaller volume leads also to an overall increase of the
exchange and self-dipolar energy, and thus a shift of the fundamental
mode with increasing energy. We have checked that the position of the
peak is compatible with the magnetic parameters shown is Table I
assuming that the center of the probe is approached from 3.0 to 1.5
$\mu$m when moved from the largest to the smallest disk.

\begin{figure}[]
  \includegraphics[width=0.9\columnwidth]{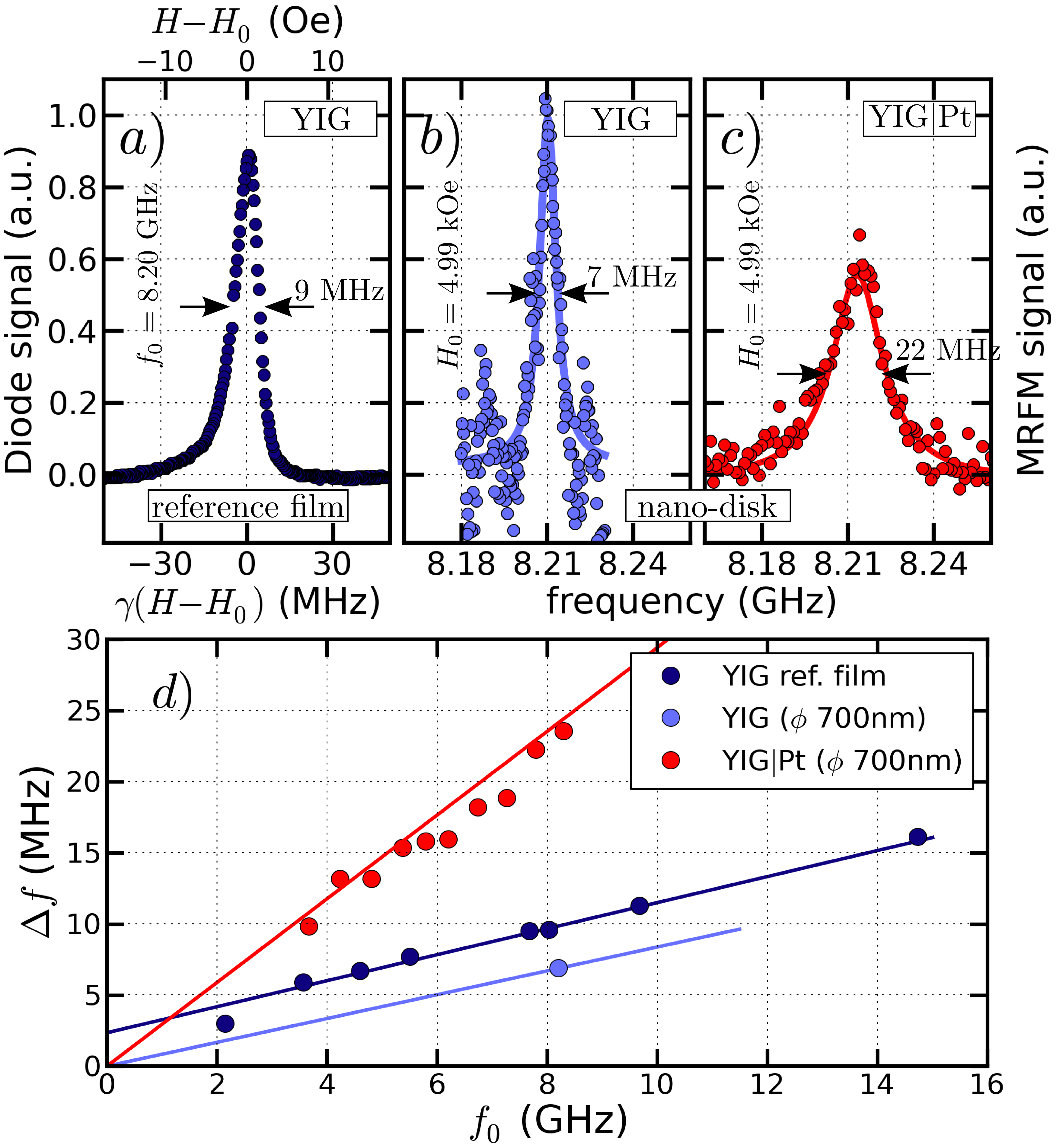}
  \caption{(a) Low-power spectrum of the reference YIG film recorded
    at $f_0=8.2$~GHz. Lineshape of the uniform mode measured at
    $H_0=4.99$~kOe on the 700~nm disk bare (b, blue circles) and with
    (c, red circles) Pt. (d) Dependence of the linewidth on the
    resonance frequency.}
  \label{fig:lwdetail}
\end{figure}

From the comparative measurements presented in
Fig.\,\ref{fig:compspec}, we see that the peak position is not altered by the addition of the Pt layer ontop of the YIG disk whereas its linewidth is clearly increased.  To emphasize this result,
Fig.\,\ref{fig:lwdetail} shows the linear f-MRFM spectra of the 700~nm
bare YIG (b, blue dots) and of the YIG{\textbar}Pt (c, red dots)
disks. For comparison the spectrum measured on the extended YIG thin
film is shown in (a, black). It was recorded with a transmission line
Au-antenna of 500~$\mu$m width and an in-plane external field oriented
perpendicularly to the exciting microwave field, similarly to the
configuration used in Ref.\cite{hahn13}.

A first striking result of Fig.\,\ref{fig:lwdetail} is obtained by
comparing the spectra measured in extended film (a) and the
nanostructure (b) for the bare YIG: the nanopatterning \emph{improves}
the linewidth \cite{loubens07}. One can observe that, while the
lineshape of the resonance has a Lorentzian shape in the nanostruture
(continuous line), the peak shape is asymetric in the YIG film. We
attribute it to inhomogeneous broadening. This is confirmed by
performing experiments on different slabs of the reference film, which
actually yield different lineshapes (not shown). The usual method of
separating the homogeneous contribution from the inhomogeneous one is
to study the frequency dependence of the half-linewidth $\Delta
f/2$. The slope gives $\alpha$, the Gilbert damping constant, while
the zero frequency intercept gives the inhomogeneous contribution. In
Fig.\,\ref{fig:lwdetail}(d), we have thus plotted the full-linewidth
$\Delta f$ of the extended reference film at different frequencies
using the same color as in Fig.\,\ref{fig:lwdetail}(a). The linewidth
varies almost linearly with the frequency. We extract from the slope
the damping $4 \cdot$ 10$^{-4}$, while the intercept at zero frequency
indicates the amount of inhomogeneity of the resonance: $\Delta
f$=2.5~MHz (or $\Delta H$=1~Oe).

On the same Fig.\,\ref{fig:lwdetail}(d), we show using red dots the
frequency dependence of the linewidth of the YIG{\textbar}Pt
nanostructure. The fit of the slope yields
$\alpha_\text{YIG{\textbar}Pt} =13 \cdot 10^{-4}$. The striking
feature is that a linear fit now intercepts with the origin of
coordinates. It means that the linewidth measured in the
nanostructures directly yields the homogeneous contribution. 
For the bare YIG nanodisk it was only possible to reliably extract the
linear linewidth at one point at 8.2 GHz (blue dot). Interestingly
enough the slope of the straight line from this point to the origin is
exactly that of the line fitted to the extended film data. We
speculate by analogy to the Pt/YIG case that the linewidth in the YIG
nanostructure is purely homogeneous in nature, while the intrinsic
part of the damping has been unaffected by the lithographic process.

In Fig.\,\ref{fig:lwdetail}, we also see that the linewidth of the
700~nm YIG{\textbar}Pt disk is 22~MHz (c) \textit{i.e.} about three
times wider than that of the 700~nm YIG disk, which is 7~MHz (b). The
influence of an adjacent Pt layer is shown to increase the damping
threefold through spin-pumping effect \cite{tserkovnyak02}.  The
characteristic parameter for the efficiency of spin transfer across
the interface and the accompanying increase of damping in YIG is the
spin mixing conductance $G_{\uparrow \downarrow}$.  One can directly
evaluate the increase of damping induced by the presence of the Pt
layer. As we find $\alpha_\text{YIG}=4 \cdot 10^{-4}$ and
$\alpha_\text{YIG{\textbar}Pt}=13 \cdot 10^{-4}$, we deduce a large
spin-pumping contribution $\alpha_{sp}= 9 \cdot 10^{-4}$ that adds to
the intrinsic damping : $\alpha_\text{YIG{\textbar}Pt}=
\alpha_\text{YIG}+\alpha_\text{sp}$. From this, one can calculate the
spin mixing conductance according to \cite{heinrich11}:
\begin{equation}
\label{eq:mix}
G_{\uparrow \downarrow} = \alpha_\text{sp} \frac{4 \pi M_S t}{g \mu_B} G_0 \, ,
\end{equation}
\vspace{0.1cm}
where $t = 20$~nm is the YIG thickness, $g$ the electron Land\'e
factor, $\mu_B$ the Bohr magneton, and $G_0=2e^2/h$ the quantum of
conductance. This corresponds to: $G_{\uparrow \downarrow}=1.55 \cdot
10^{14} \Omega^{-1}\text{m}^{-2}$ for our YIG{\textbar}Pt interface.

In summary, we have conducted a study of spin wave spectra of
individual submicron YIG and YIG{\textbar}Pt disks. We find that the
litography process does not broaden the linewidth and on the contrary,
the linewitdh decreases compared to the extended film. The influence
of an adjacent Pt layer on the YIG through the spin pumping effect is
investigated and quantified to increase the damping 3 fold. As a non
zero spin mixing conductance is determined, these experiments pave the
way for observation of inverse spin Hall effects in a YIG{\textbar}Pt
nanodisk accessing its individual spin wave modes. Most importantly,
thanks to the small volume and purely intrinsic damping of the YIG
sample, we will address in future studies the influence of direct spin
Hall effect on the linewidth of these YIG nano-disks as it was
demonstrated in all-metallic NiFe/Pt dots \cite{demidov12}.

\begin{acknowledgments}
  This research was supported by the French Grants Trinidad (ASTRID
  2012 program) and by the RTRA Triangle de la Physique grant
  Spinoscopy. We also acknowledge usefull contributions from
  C. Deranlot, A.H. Molpeceres and R. Lebourgeois.
\end{acknowledgments}


%

\end{document}